\documentclass[journal,twoside,web]{ieeecolor}

\usepackage{generic}
\usepackage[utf8]{inputenc}
\usepackage{cite}
\usepackage[utf8]{inputenc}
\usepackage{cite}
\usepackage{qtree}
\usepackage{tikz}
\usepackage{amsmath,amssymb,amsfonts}
\usepackage{eufrak}
\usepackage{dsfont}
\usepackage{steinmetz}
\usepackage{algorithmicx}
\usepackage{algorithm}
\usepackage{algpseudocode}
\usepackage{graphicx}
\usepackage{amsmath,amssymb}
\usepackage{textcomp}
\usepackage{algpseudocode} 
\usepackage{mathtools}
\usepackage{nicefrac}
\usepackage{txfonts}
\usepackage{tikz}
\usepackage{dsfont}
\usepackage{graphicx}
\usepackage{amsmath,amssymb}
\usepackage{xcolor}
\usepackage{mathtools}
\usepackage{nicefrac}
\usepackage{txfonts}
\usepackage{todonotes}
\usepackage{cleveref}
\usepackage{subfig}
\usepackage[font=small]{caption}

\newcommand{\R}{\mathbb{R}}

\newcommand{\E}{\mathbb{E}}

\newcommand{\T}{^{\mbox\tiny\mathsf{T}}}
\newcommand{\tr}{{\rm{tr}}}

\newcommand{\K}{\mathcal{K}}
\newcommand{\Q}{\mathcal{Q}}
\newcommand{\Qw}{\mathcal{Q}^w}

\newcommand{\Kd}{\K_{\rm EDMD}}

\newcommand{\Kdt}{\tilde{\K}_{\rm EDMD}}

\renewcommand{\baselinestretch}{0.965}
\allowdisplaybreaks

\DeclareMathOperator{\argmin}{\arg\!\min}

\def\mf{\mathbf}
\def\mb{\mathbb}
\def\mc{\mathcal}
\def\beq{\begin{equation*}}
\def\eeq{\end{equation*}}
\def\bql{\begin{equation}}
\def\eql{\end{equation}}
\def\bqn{\begin{eqnarray*}}
\def\eqn{\end{eqnarray*}}
\def\bnl{\begin{eqnarray}}
\def\enl{\end{eqnarray}}
\def\bna{\bql\begin{array}{rcl}}
\def\ena{\end{array}\eql}
\def\bnn{\beq\begin{array}{rcl}}
\def\enn{\end{array}\eeq}
\def\bma{\begin{bmatrix}}
\def\ema{\end{bmatrix}}
\def\bmx{\begin{matrix}}
\def\emx{\end{matrix}}
\def\ben{\begin{enumerate}}
\def\een{\end{enumerate}}
\def\bit{\begin{itemize}}
\def\eit{\end{itemize}}
\def\bei{\begin{itemize}}
\def\eei{\end{itemize}}
\def\bet{\begin{tabular}}
\def\eet{\end{tabular}}

\newtheorem{thm}{Theorem}

\newtheorem{pr}{ Proposition}
\newtheorem{rem}{ Remark}

\newtheorem{asm}{Assumption}

\newcommand{\DQEDMD}{\texttt{DQ-EDMD}}

\title{On the Effect of Quantization on Extended Dynamic Mode Decomposition}
\author{ Dipankar Maity and Debdipta Goswami
\thanks{Both the authors contributed equally to this work.}%
\thanks{D. Maity is with the Department of Electrical and Computer Engineering and an affiliated faculty of the North Carolina Battery Complexity, Autonomous Vehicle, and Electrification Research Center (BATT CAVE), University of North Carolina at Charlotte,  NC, 28223, USA.
Email: {\tt {dmaity@charlotte.edu}}
}
\thanks{
D. Goswami is with the Department of Mechanical and Aerospace Engineering, The Ohio State University, Columbus,
OH, 43210, USA. Email:{ \tt goswami.78@osu.edu }
}
}

\begin{document}
\maketitle
\thispagestyle{empty}

\begin{abstract}

Extended Dynamic Mode Decomposition (EDMD)  is a widely used data-driven algorithm for estimating the Koopman Operator. EDMD extends Dynamic Mode Decomposition (DMD) by lifting the snapshot data using nonlinear dictionary functions before performing the estimation.
    This letter investigates how the estimation process is affected when the data is quantized.
    Specifically, we examine the fundamental connection between estimates of the operator obtained from unquantized data and those from quantized data via EDMD.
    Furthermore, using the law of large numbers, we demonstrate that, under a \textit{large data regime}, the quantized estimate can be considered a regularized version of the unquantized estimate. 
    %
    %
    We also explore the relationship between the two estimates in the \textit{finite data regime}.
    We further analyze the effect of nonlinear lifting functions on this regularization due to quantization.
    The theory is validated through repeated numerical experiments conducted on two different dynamical systems.
    
\end{abstract}

\begin{IEEEkeywords}
System identification, EDMD, Quantization.
\end{IEEEkeywords}

\section{Introduction} 
\IEEEPARstart{S}{ystem identification} is an essential component in the applications of controls and dynamical systems  involving unknown or partially known dynamics.
Extended Dynamic Mode Decomposition (EDMD) \cite{Williams2015} is a Koopman operator theory \cite{Koopman1931} based data-driven system identification algorithm that estimates a finite-dimensional representation of the Koopman operator.
EDMD solves a least-square optimization problem  for estimating the Koopman operator \cite{Williams2015} using data snapshots from the dynamical system.
It is well-understood that the quality of the Koopman operator estimate improves/degrades with an increase/decrease in the amount of data, as expected \cite{Arbabi2017, Hirsh2020, Lee2024}.
On the other hand, it is not clear how the quality of the data affects the estimation process, especially when the data undergoes a quantization process.

Existing work in EDMD and data-driven system identification typically assumes that these algorithms are implemented on systems with ample resources to handle large datasets generated from snapshots of the dynamical system. 
However, applying these data-intensive algorithms to resource-limited systems, such as low-powered, lightweight robotic applications \cite{Folkestad2022, Cleary2020}, may require quantization to meet hardware and other resource constraints. 
In fact, quantization naturally arises under communication and computation constraints, making it a common practice in networked control systems, multi-agent systems, and cyber-physical systems in general.

Quantization can have severe consequences on control systems, to the extent that a stabilizable system becomes destabilized if the quantization word length falls below a certain threshold \cite{nair2004stabilizability}.
Since system identification is often the first step in controlling unknown systems, the effects of quantization on the identification process will, in turn, affect controllers, state estimators, and ultimately the overall performance of the system.
The choice of quantizer is also of particular significance, as it can affect the system's performance \cite{maity2021optimal, maity2023optimal}.

In this letter, we study the effects of \textit{dither quantization} \cite{gray1993dithered} ---a highly effective and commonly used quantization method in controls, communications, and signal processing--- on EDMD by extending our prior work \cite{maity2024effect}. 
To the best of our knowledge, \cite{maity2024effect} is the first work to investigate the effect of quantization on Dynamic Mode Decomposition (DMD) and develops and analyzes the \textit{Dither Quantized DMD} method.
In contrast to DMD, the EDMD method involves lifting the state-data via a dictionary of observable functions \cite{Williams2015}, and consequently,  those lifting functions (e.g., radial basis functions, Legendre polynomials) may further amplify the effects of quantization. This letter discusses the impact of quantization and examines the role of lifting functions in influencing the resulting effects. 
It offers insights into which dictionary may be more favorable when quantization is a factor.


\color{black}
The rest of the paper is organized as follows: \Cref{sec:background} provides the necessary background materials on Koopman Operator theory, Extended Dynamic Mode Decomposition, and Dither Quantization. 
We define our problem statement in \Cref{sec:ProblemStatement} and  analyze the dither quantized extended dynamic mode decomposition (\DQEDMD) in \Cref{sec:QuantizedDMD}, demonstrating the connection between the solution obtained from $\DQEDMD$  and unquantized EDMD. 
We discuss our observations from implementing $\DQEDMD$ on two dynamical systems in \Cref{sec:Validation} and we provide some conclusions in \Cref{sec:conclusions}.

\textit{Notations:} Set of non-negative integers are denoted by $\mb{N}_0$. 
$(\cdot)^{\dagger}$ and $(\cdot)^\top$ denote the Moore--Penrose inverse and transpose of a matrix, respectively. 
$\|\cdot\|$ denotes a norm, where we use Euclidean norm for vectors and Frobenius norms for matrices. 
The Big-O notation is denoted by $O(\cdot)$.

\section{Background} \label{sec:background}
\subsection{Koopman Operator Theory}
Consider a discrete-time dynamical system on an $n$-dimensional compact manifold $\mathcal{M}$, evolving according to the flow-map ${f}:\mc{M}\mapsto \mc{M}$ as follows: 
\begin{equation} \label{Eq: Dynamics}
    {x}_{t+1} = {f}({x}_{t}),\quad {x}_t\in\mc{M},\quad t\in\mb{N}_0.
\end{equation}
Let $\mc{F}$ be a Banach space of complex-valued observables  $\varphi:\mc{M}\rightarrow \mb{C}$. The discrete-time \emph{Koopman operator} $\mc{K}:\mc{F}\rightarrow \mc{F}$ is defined as
\begin{equation}
    \mc{K}\circ\varphi(\cdot) = \varphi \circ {f}(\cdot),\quad \text{with}~~\varphi({x}_{t+1})=\mc{K}\varphi({x}_{t}),
\end{equation}
where $\mc{K}$ is infinite-dimensional, and linear over its argument. The scalar observables $\varphi$ are referred to as the Koopman observables.

A Koopman eigenfunction $\phi_i$ is a special  observable that satisfies $(\mc{K}\phi_i)(\cdot)=\lambda_i \phi_i(\cdot)$, for some eigenvalue $\lambda_i \in \mb{C}$. 
Considering the Koopman eigenfunctions (i.e., $\{\phi_i\}_{i \in \mb{N}}$) span the Koopman observables,  any {vector valued observable ${g}=[\varphi_1,~\varphi_2,~\ldots,~\varphi_p]^\top \in\mc{F}^p $} can be expressed as a sum of Koopman eigenfunctions ${g}(\cdot)=\sum_{i=1}^{\infty}\phi_i(\cdot){v}^{{g}}_i$, where ${v}^{{g}}_i\in\mb{R}^p$, for $i=1,2,\ldots,$ are called the \emph{Koopman modes} of the observable $g(\cdot)$. This modal decomposition provides the growth/decay rate $|\lambda_i|$ and frequency $\angle{\lambda_i}$ of different Koopman modes via its time evolution:
\begin{equation}\label{eq:koop_decomp}
    {g}({x}_t) = \sum\nolimits_{i=1}^{\infty}\lambda_i^t\phi_i({x}_0){v}^{{g}}_i.
\end{equation}
\noindent The Koopman eigenvalues ($\lambda_i$) and eigenfunctions ($\phi_i$) are properties of the dynamics only, whereas the Koopman modes ($v^i_g$) depend on the observable ($g$). 


Several methods have also been developed to compute the Koopman modal decomposition, e.g., DMD and EDMD \cite{schmid2010, Williams2015}, Ulam-Galerkin methods, and deep neural networks \cite{otto2019linearly, Yeung2019}. 
In this work, we focus on the EDMD method, which is briefly described below. 

\subsection{Approximation of Koopman Operator via EDMD}
Extended dynamic mode decomposition  is a data-driven method for approximating Koopman operator and dominant Koopman modes from a sequence of time-series data using a set of observable functions and matrix factorization. It was developed \cite{Williams2015} as a nonlinear extension of dynamic mode decomposition to extract spatio-temporal structures from intricate flows. EDMD uses a set of observables or dictionary functions  $\varphi(\cdot) = [\varphi^1(\cdot),\ldots,\varphi^N(\cdot)]^\top: \mc{M} \mapsto \mb{C}^N$ to lift the state $x$ to an $N$-dimensional latent space. EDMD requires a pair of snapshot data matrices in order to generate a linear model approximating the desired dynamical system. They are created by sampling the  state variables $x\in\mb\R^n$ at a sequence of time instants (snapshots) and concatenating them to form snapshot matrices $\mf{X} \in \mb{R}^{n\times T}$ and $\mf{X}' \in \mb{R}^{n\times T}$, where $\mf{X}'$ is one time snapshot ahead of the original snapshot matrix $\mf{X}$, i.e.,
\begin{align}
    \mf{X} = 
    \begin{bmatrix}
        {x}_0 & ... & {x}_{T-1}\\
    \end{bmatrix},
    \quad
    \mf{X}' = 
    \begin{bmatrix}\label{eq:dmd_snap_mat}
        {x}_{1} &  ... & {x}_{T}\\
    \end{bmatrix}.
\end{align}  Now, define lifted snapshot matrices $ \Phi,  \Phi' \in \R^{N \times T}$ such that
\begin{align} \label{eq:dataMatrix}
\begin{split} 
    \mf{\Phi} &= \begin{bmatrix}
         \varphi(x_0) ~  \varphi(x_1) ~ \hdots ~ \varphi(x_{T-1}) 
    \end{bmatrix},  \\
    \mf{\Phi}' &= \begin{bmatrix}
         \varphi(x_1) ~  \varphi(x_2) ~ \hdots ~ \varphi(x_{T}) 
    \end{bmatrix} .
\end{split}    
\end{align} 

The EDMD algorithm aims to find the best linear operator $\Kd$ that relates the two lifted snapshot matrices $\mf{\Phi}$ and $\mf{\Phi}'$ in a least-square sense, i.e.,
\begin{align}\label{eq:dmd_linear_fit}
    \mf{\Phi}' &\approx {\Kd}\mf{\Phi},
\end{align}
where $\Kd =\argmin_{A\in\R^{N\times N}} \frac{1}{T} \|\mf{\Phi}' - {A}\mf{\Phi}\|^2$. An approximate linear map $C$ mapping from $\operatorname{span}\{\varphi^1(\cdot),\ldots,\varphi(\cdot)\}$ to $\R^n$ is found by solving $\mc{C} =\argmin_{C\in\R^{n\times N}} \frac{1}{T} \|\mf{X} - {C}\mf{\Phi}\|^2$.

The $\Kd$ matrix represents the Koopman operator in the newly mapped linear space of finite-dimensional observables.
EDMD is performed by computing the pseudo-inverse of $\mf{\Phi}\mf{\Phi}^\top$, and then it is used for the prediction of $x_t$:
\bnl\label{Eq: DMD}
\begin{split}
 \Kd &= \mf{\Phi}\mf{\Phi}^\top\left(\mf{\Phi}'\mf{\Phi}^\top\right)^{\dagger},\\
 \hat{x}_t &= C(\Kd)^t \varphi(x_0).
 \end{split}
 \enl

\begin{rem}
    If $\phi(\cdot)$ is an identity map, i.e., $\mf{\Phi}=\mf{X}$ and $\mf{\Phi}'=\mf{X}'$, EDMD reduces to DMD.
\end{rem}

\subsection{Dither Quantization} \label{Sec:quant}

A quantizer $q :(u_{\min},u_{\max})\subseteq \R \to \{0,\ldots, (2^b-1)\}$ is a function that maps any $x\in (u_{\min},u_{\max}) \subset \R $ to a $b$-bit binary word.
A uniform quantizer takes the form 
\begin{align*}
    q(x)=\left\lfloor\frac{x-u_{\min}}{\epsilon} \right\rfloor,
\end{align*}
where 
\begin{align} \label{eq:quantizationResolution}
    \epsilon=\frac{u_{\max}-u_{\min}}{2^b}
\end{align}
denotes the quantization resolution.
Although $q(\cdot)$ is defined on the interval $(u_{\min},u_{\max})$, one may extend the definition of $q(\cdot)$ on the entire real line:
\begin{align*}
    \bar{q}(x)=\begin{cases}
    q(x),\quad &x\in (u_{\min},u_{\max}),\\
    0, & x\le u_{\min},\\
    2^b-1, & x\ge  u_{\max},
    \end{cases}
\end{align*}
where $\bar q(\cdot)$ is the extended version of $q(\cdot)$. 
The region outside the interval $[u_{\min},u_{\max}]$ is referred to as the \textit{saturation} region of the quantizer $\bar{q}$.

The decoding of a mid-point uniform quantizer is given by
\begin{align} \label{eq:decoding}
    \Q(x) =\epsilon q(x) + u_{\min} + \frac{\epsilon}{2}.
\end{align}
The quantization error is defined to be $e(x) = \mathcal{Q}(x) - x$. 
For all $x\in (u_{\min}, u_{\max})$, we have $|e(x)| \le \frac{\epsilon}{2} $.
The distribution of the quatization error plays an important role in analyzing the performance of a system employing quantization. 
The distribution of this error is correlated with the distribution of the source signal $x$.
This correlation may result in poor performance, besides making the analysis of such systems complicated. 
It has been well-established that \textit{dither quantization} leads to a better performance, as demonstrated in the very first work on TV communication \cite{roberts1962picture} as well as in applications to controls \cite{stanford1960linearization}. 
Since then, a significant amount of research has been devoted in \textit{dither} quantization. 

\textit{Dither} quantization prescribes adding a noise $w$ to the source signal $x$ prior to quantization and subtract that noise during decoding \cite{gray1993dithered}, which yields the decoded signal to be
\begin{align} \label{eq:decoding_dither}
    \tilde{x} = \Q(x+w) - w,
\end{align}
where $\Q(\cdot)$ is defined in \eqref{eq:decoding}.
Thus, the quantization error under the dithering scheme becomes 
\begin{align}
    e(x) = \mathcal{Q}(x+w) - w - x.
\end{align}
Under certain assumptions on the distribution of $w$, it can be shown that this new error $e(x)$ is distributionally independent of the source $x$. 
Furthermore, this error can be shown to have a uniform distribution in $[-\frac{\epsilon}{2}, \frac{\epsilon}{2}]$. 
A typical choice of $w$ is to consider an uniformly distributed random variable with support $[-\frac{\epsilon}{2}, \frac{\epsilon}{2}]$, which satisfies all the necessary and sufficient conditions to ensure that $e $ is independent of $x$ and uniformly distributed in $[-\frac{\epsilon}{2}, \frac{\epsilon}{2}]$; see \cite{gray1993dithered}. 
Throughout this work, we will consider this \textit{dither} quantization scheme. 

When $x$ is a vector, we perform the quantization and the decoding component-wise. 
That is, for $x\in \R^n$, we generate a dither vector $w\in \R^n$ where the components of $w$ are independent random variables. 
Consequently, we define the quantization error of the $j$-th component as $e^j(x) = \Q(x^j + w^j) - w^j - x^j$. Under the \textit{dither} assumption, $e^j(x)$ is independent of $x$ and $e^k(x)$ for all $k\ne j$.

For quantizing a time-varying vector-valued process $\{x_t\}_{t\ge 0}$, we will consider a time-varying vector-valued i.i.d process $\{w_t\}_{t\ge 0}$ as the \textit{dither} signal. 
Furthermore, in the subsequent sections we will use $\Qw(x)$ as a shorthand notation for $\Q(x+w)$.

\section{Problem Statement} \label{sec:ProblemStatement}



%
In this work, our objective is to understand the effects of dither quantization on the estimated Koopman operator. 
To that end, we assume that the observables are computed based on decoded quantized data $\tilde{x}$, where $\tilde{x}$ is defined in \eqref{eq:decoding_dither}.
That is, the data pertaining to the $i$-th observable at time $t$ is $\varphi^i (\tilde{x}_t)$, whereas, in the unquantized case, that data is $\varphi^i(x_t)$. 
Consequently, at time $t$, the available data is
\begin{align} \label{eq:tildeObservables}
    \bar \varphi (x_t) \triangleq \begin{bmatrix}
        \varphi^1(\tilde x_t), \cdots,
        \varphi^N(\tilde x_t)
    \end{bmatrix}\T, 
\end{align}
where $\tilde{x}_t = \Qw(x_t) - w_t$ is the decoded state measurement, and $w_t$ is the dither signal used in the quantization of the state at time $t$. 

Let $\Kdt$ denote the estimate of the Koopman operator  obtained from the quantized data.
That is, 
\begin{align} \label{eq:EDMD_quantized}
      \Kdt = \argmin_{A \in \R^{N \times N}} \frac{1}{T}  \| \bar \Phi' - A \bar \Phi\|^2,
\end{align}
where
\begin{align*}
    \bar \Phi &= \begin{bmatrix}
        \bar \varphi(x_0) ~ \bar \varphi(x_1) ~ \hdots ~ \bar \varphi(x_{T-1}) 
    \end{bmatrix},~
    \bar \Phi' = \begin{bmatrix}
        \bar \varphi(x_1) ~ \bar \varphi(x_2) ~ \hdots ~ \bar \varphi(x_{T}) 
    \end{bmatrix} ,
\end{align*}
and where $\bar{\varphi}(\cdot)$ is defined in \eqref{eq:tildeObservables}.
On the other hand, the estimate obtained from the unquantized data is
\begin{align} \label{eq:EDMD}
    \Kd = \argmin_{A \in \R^{N \times N}}  \frac{1}{T}\| \Phi' - A \Phi\|^2, 
\end{align}
where $ \Phi,  \Phi' \in \R^{N \times T}$ are the data matrices defined in \eqref{eq:dataMatrix}.
Under the dither quantization scheme, the optimization in \eqref{eq:EDMD_quantized} is referred to as the $\DQEDMD$ problem.

\begin{rem}
    Due to the injected noise in the dither quantization, $\Kd$ is a random matrix whose realization is coupled with the realization of the dithering noise.
    Therefore, when we say $\Kdt \to \K^*$, then such convergence should be interpreted in the sense of covergence of random variables. 
\end{rem}

Having obtained the matrices $\Kd$ and $\Kdt$, we may predict the state using \eqref{Eq: DMD}.
In this letter, we investigate the normalized errors both in the estimation of the Koopman operator and in the prediction of system's state.  
That is, we quantify how $\frac{\| \Kd - \Kdt\|}{\|\Kd\|}$ and $\frac{1}{T}\sum_{t=0}^{T-1} \frac{\|\hat{x}_t - x_t \|}{\|x_t\|}$ change as we vary the word length for the quantization.

In addition to quantifying the degradation due to quantization using the aforementioned metrics, we are also interested in developing a framework where one may obtain an improved estimate, $\Kdt^*$, that is closer to $\Kd$ than $\Kdt$ is. 
In this work we discuss such a potential method for the large data regime (i.e., when $T\to \infty$).

\section{EDMD with Quantized Data} \label{sec:QuantizedDMD}

In this section, we study EDMD under both large (i.e, $T\to \infty$) and finite (i.e., $T<\infty$)  data regimes. 
In the large data regime we show that $\Kdt$ and $\Kd$ are connected via a regularized optimization problem.
In the finite data regime, we show that the difference between between $\Kdt$ and $\Kd$ is $O(\epsilon)$, with $\epsilon$ being the quantization resolution. 

\subsection{Large Data regime}





Define the one-step least-square residual {\small $r:\mc{M}\times\mc{M}\rightarrow \R_+$} such that {\small$r(x_{t+1},x_t)\triangleq \|\varphi(x_{t+1})-A\varphi(x_t)\|^2$}. 
Therefore, we may write 
\begin{align*}
    \lim_{T\to \infty} \frac{1}{T} \|\Phi' - A\Phi\|^2 = \lim_{T\to \infty} \frac{1}{T} \sum\nolimits_{t=0}^{T-1} r(x_{t+1},x_t)
\end{align*}
For our analysis, we make the following assumptions. 
\begin{asm} \label{assm:boundedR}
    There exists $\mathcal{A} \subseteq \R^{N\times N}$ and $c_r > 0$ such that $r(x_{t+1}, x_t)< c_r$ for all $t$ when $A\in \mathcal{A}$.
\end{asm}
\begin{asm} \label{assm:AbsConvTaylor}
     $r(\cdot,\cdot)$ has an absolutely convergent Taylor series. 
\end{asm}
\begin{asm} \label{assm:BoundedPhi_derivative}
    There exists $c_\varphi > 0$  such that $\|\nabla \varphi_i(x) \| \le c_\varphi$ for all $x \in \R^n$ and $i=\{1,\ldots, N\}$.
\end{asm}
\Cref{assm:boundedR} is necessary and sufficient to ensure that the unquantized EDMD under the large data regime is a well-posed problem.
Notice that the set $\mathcal{A}$ in \Cref{assm:boundedR} is equivalent to the set 
$     \mathcal{A} = \{A  : \lim_{T\to \infty} \frac{1}{T} \|\Phi' - A\Phi\|^2 <+\infty \}.$ 
\Cref{assm:AbsConvTaylor} is used in \Cref{thm:equivalence} and \Cref{assm:BoundedPhi_derivative} is used in both Theorems~\ref{thm:equivalence} and \ref{thm:K_epsilon}.

\begin{thm}[Large data regime result]
\label{thm:equivalence}
    As $T \to \infty$ and $|\epsilon|<1$, $\Kdt$ converges almost surely to the solution of the following regularized EDMD   \vspace{ -1mm}
    \begin{align} \label{eq:equivalence}
    \begin{split}
        \min_{A \in \R^{N \times N}} \limsup_{T\to \infty}\frac{1}{T}  \| \Phi' - A \Phi\|^2 + \tr(A\beta(\epsilon)) + \tr(A\T A \Gamma(\epsilon)),
        \end{split}
    \end{align} \vspace{-1mm}
    where $\beta(\epsilon)$ and $\Gamma(\epsilon)$ are $O(\epsilon^2)$, and $O(\cdot)$ is the Big-O notation.
\end{thm}

\begin{proof}
    The proof is presented in \Cref{AP:thm:equivalence}.
\end{proof}
\vspace{2 mm}

Theorem~\ref{thm:equivalence} implies that  $\Kdt$ can be interpreted as a solution to a regularized EDMD problem, where the regularization parameter depends on the quantization resolution  $\epsilon$ and some matrices $\beta$ and $\Gamma$ which depend on the Taylor series coefficients of the residual $r(\cdot,\cdot)$.  
A consequence of \Cref{thm:equivalence} is that the solution {\small$\Kdt$} converges to {\small$\Kd$} almost surely as $\epsilon$ approaches to $0$. 
Recall from \eqref{eq:quantizationResolution} that the quantization resolution $\epsilon$ is coupled with the qunatization word length $b$. 
Thus, as $b$ increases, we obtain {\small$\Kdt \to \Kd$} almost surely, as one would expect.

\begin{rem}
    \Cref{thm:equivalence} shows the fundamental connection between {\small$\Kd$} and {\small$\Kdt$}. 
    Note that the relationship \eqref{eq:equivalence} holds because the quantization noises are i.i.d., which is due to the fact that dither quantization is being used. 
    A similar conclusion may not hold for other forms of quantization. 
\end{rem}

\Cref{thm:equivalence} not only helps in identifying the relationship between $\Kd$ and $\Kdt$, but also provides a convenient framework to potentially recover $\Kd$ from the quantized data, as discussed next. 

\subsection{Regularized $\DQEDMD$}

\Cref{thm:equivalence} demonstrates that {\small$\frac{1}{T}  \| \bar \Phi' - A \bar \Phi\|^2$} almost surely converges to {\small$\frac{1}{T}  \| \Phi' - A \Phi\|^2 + \tr(A\beta(\epsilon)) + \tr(A\T A \Gamma(\epsilon)) + \text{constant}$}, as ${T\to \infty}$.
Alternatively, one may state that {\small$\frac{1}{T}  \| \bar \Phi' - A \bar \Phi\|^2 - + \tr(A\beta(\epsilon)) - \tr(A\T A \Gamma(\epsilon))$} almost surely converges to {\small$\frac{1}{T}  \| \Phi' - A \Phi\|^2$} + \text{constant}. 
Therefore, one may further claim that 
\begin{align} \label{eq:regularized_EDMD}
    \argmin_{A} \limsup_{T\to \infty} \frac{1}{T}  \| \bar \Phi' - A \bar \Phi\|^2 -  \tr(A\beta(\epsilon)) - \tr(A\T A \Gamma(\epsilon)) \nonumber \\
   = \argmin_{A} \limsup_{T\to \infty} \frac{1}{T}  \|  \Phi' - A \Phi\|^2 = \Kd.
\end{align}
In other words, $\Kd$ can be recovered from quantized data by solving the regularized $\DQEDMD$ problem defined in \eqref{eq:regularized_EDMD}, where $\beta(\epsilon), \Gamma(\epsilon)$ are the regularization parameters. 
The challenge in recovering {\small$\Kd$} from \eqref{eq:regularized_EDMD} is that the exact expressions of the regularization parameters are not easy to obtain. 
One potential approach would be to approximate these quantities by $\hat{\beta}(\epsilon)$ and $\hat{\Gamma}(\epsilon)$. 
Such approximation is beyond of the scope of this letter and will be addressed in future works. 
In this letter, we state a special case where one may compute these quantities. 

\begin{pr}
    For the case of DMD, where the observables are identity mappings,  $\Gamma(\epsilon) = \frac{\epsilon^2}{12}I$ and $\beta(\epsilon) = 0$. 
\end{pr}

\begin{proof}
    A proof of this can be found in \cite{maity2024effect}.
\end{proof}

\subsection{Finite Data Regime}



\begin{thm}[Finite data regime result] \label{thm:K_epsilon}
    Let $\Phi$ and $\bar \Phi$ be of full row rank. Then, $\exists \ \K_\epsilon$ such that $\|\K_\epsilon\| = O(\epsilon)$ and 
    \begin{align}
    \Kdt = \Kd + \K_\epsilon.
    \end{align}
\end{thm}

\begin{proof}
    The closed form solution to the EDMD problem in \eqref{eq:EDMD_quantized} with quantized data  is 
\begin{align} \label{eq:KDt_solution}
    \Kdt = \bar\Phi' \bar \Phi^\top \big( \bar \Phi \bar \Phi^\top\big)^{-1}, 
\end{align}
whereas that for the unquantized EDMD is {\small$\Kd = \Phi'\Phi\T(\Phi\Phi\T)^{-1}$}.

Due to mean-value theorem, we may write 
\begin{align} \label{eq:mean-value-theorem}
     \varphi^i(\tilde{x}_t) & = \varphi^i(x_t) + \underbrace{e_t\T \nabla \varphi^i(x_t + \alpha^i_t e_t)}_{ \triangleq \delta^i_{t}} 
\end{align}
for some $\alpha^i_t \in [0,1]$. 
Consequently, $\bar \Phi = \Phi + \Phi_\epsilon$, where the $ij$-th element of $\Phi_\epsilon$ is the {\small$\delta^i_j$} defined in \eqref{eq:mean-value-theorem}. 
Notice that $|\delta^i_t| \le \frac{nc_\varphi}{2}\epsilon$  since $\| \nabla \varphi^i (x)\| \le c_\varphi$ for all $x$ due to \Cref{assm:BoundedPhi_derivative}, and $\|e_t\| \le \frac{\sqrt{n}}{2}\epsilon$ due to the quantization process.
Thus, $\| \Phi_\epsilon\| = O(\epsilon)$.

Substituting $\bar \Phi = \Phi + \Phi_\epsilon$ and $\bar \Phi' = \Phi' + \Phi'_\epsilon$  in \eqref{eq:KDt_solution} followed by some simplifications yields
\begin{align*}
    \Kdt & = \Kd - \Kd\big(  \Phi  \Phi^\top\Psi_\epsilon^{-1} + I  \big)^{-1} + \Pi_\epsilon \big( \bar \Phi \bar \Phi^\top\big)^{-1} ,
\end{align*}
where $\Psi_\epsilon = \Phi_\epsilon \Phi^\top + \Phi \Phi_\epsilon^\top + \Phi_\epsilon \Phi_\epsilon^\top$  and $\Pi_\epsilon = \Phi'_\epsilon \Phi^\top + \Phi' \Phi_\epsilon^\top + \Phi'_\epsilon \Phi_\epsilon^\top$. 
Therefore, we may write 
\begin{align*}
    \Kdt = \Kd + \K_\epsilon,
\end{align*}
where $\K_\epsilon =  \Pi_\epsilon \big( \bar \Phi \bar \Phi^\top\big)^{-1} - \Kd\big(  \Phi  \Phi^\top\Psi_\epsilon^{-1} + I  \big)^{-1}$. 
The theorem is proven once we show that $\|\K_\epsilon\| = O(\epsilon)$. 
To that end, let us note that $\|\Phi_\epsilon\| = O(\epsilon)$ implies $\|\Psi_\epsilon\| = O(\epsilon)$ and $\|\Pi_\epsilon\| = O(\epsilon)$, and therefore, $\|\K_\epsilon \| = O(\epsilon)$. 
This concludes the proof.
\end{proof}

Using \Cref{thm:K_epsilon} we may upper bound the normalized difference between $\Kd$ and $\Kdt$:
\begin{align*}
    \frac{\| \Kd - \Kdt\|}{\|\Kd\|} \le \left( \|\Psi_\epsilon\| +\frac{\|\Pi_\epsilon\|}{\Kd} \right)\|\big( \bar \Phi \bar \Phi^\top\big)^{-1}\|.
\end{align*}
Therefore, as $\epsilon$ decreases, the upper bound on $\frac{\| \Kd - \Kdt\|}{\|\Kd\|}$ also decreases, as one would expect. 
We also observe that $\|\Psi_\epsilon\|$ and $\|\Gamma_\epsilon\|$ are proportional to the gradient bound $c_\varphi$ (c.f.~\Cref{assm:BoundedPhi_derivative}) and therefore,  lifting functions with lower $c_\varphi$ are preferable in mitigating the effects of quantization. \vspace{-0.2cm}  

\section{Numerical Examples} \label{sec:Validation}\vspace{-0.1cm}

The effect of dither quantization on data for EDMD is demonstrated on  two different systems: a simple pendulum with negative damping and Van der Pol oscillator. 
\subsection{Pendulum with negative damping}
A two dimensional oscillatory system with slight instability is considered as a first example.
The dynamics of a simple pendulum with a destabilizing term is described as:
\bnl \label{Eq: Pendulum}
\dot{x}_1 &=& x_2\nonumber\\
\dot{x}_2&=&0.01x_2-\sin x_1.
\enl
The dynamics is discretized using the fourth order Runge-Kutta method with discretization period $\Delta t = 0.01s$.
We simulate 200 trajectories over 1000 sampling periods (i.e., 10 seconds per trajectory). The initial conditions are generated randomly with uniform distribution on the unit box $[-1, 1]^2$. The lifting functions $\varphi^i$ are chosen to be the state itself (i.e., $\varphi^1=x_1$, $\varphi^2=x_2$) and 100 thin plate spline radial basis functions with centers selected randomly with uniform distribution on the unit box\footnote{Thin plate spline radial basis function with center at $x_0$ is defined by $\psi(x) = \|x - x_0\|^2\log(\|x - x_0\|)$.}, leading to a lifted state-space of dimension $N = 102$.

\begin{figure*}[t]
\centering 
\subfloat[]{\includegraphics[trim=0cm 0cm 0cm 0cm, clip=true, width=0.33\textwidth]{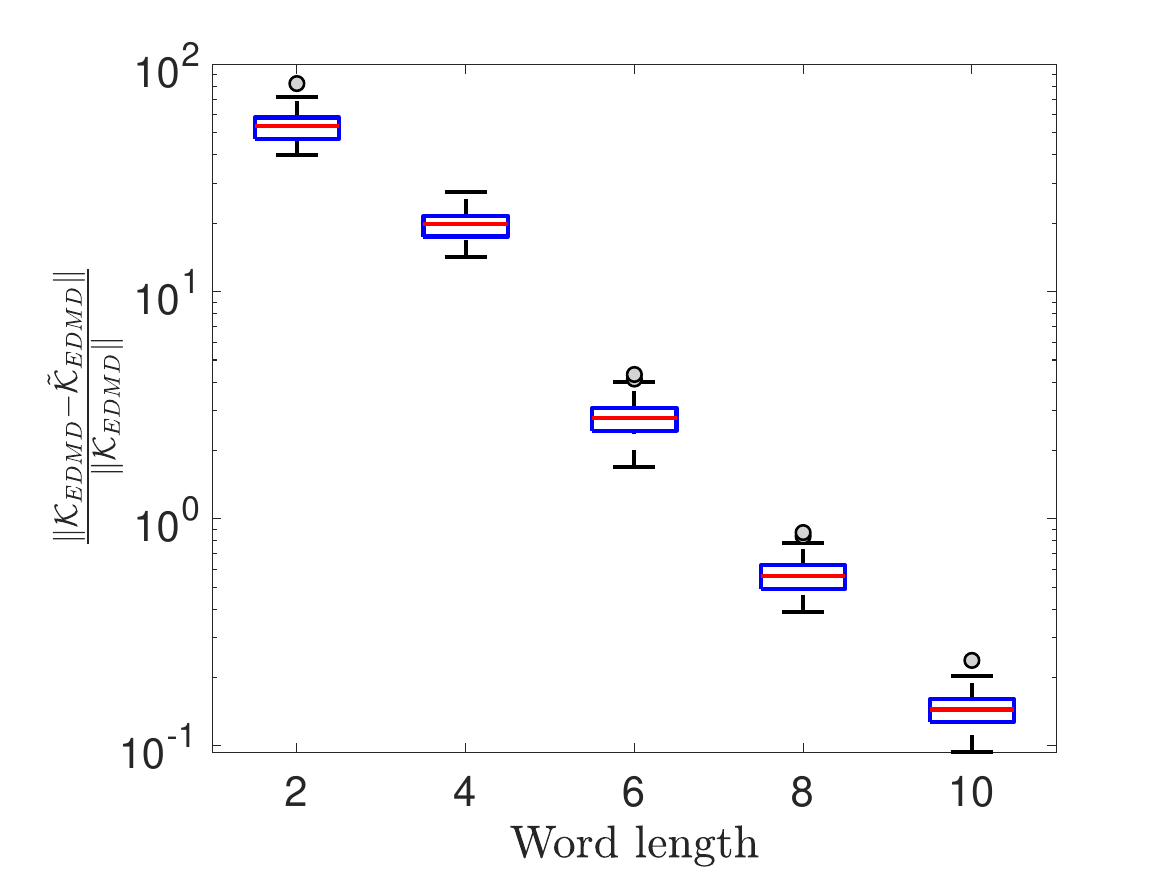}}
\subfloat[]{\includegraphics[trim=0cm 0cm 0cm 0cm, clip=true, width=0.33\textwidth]{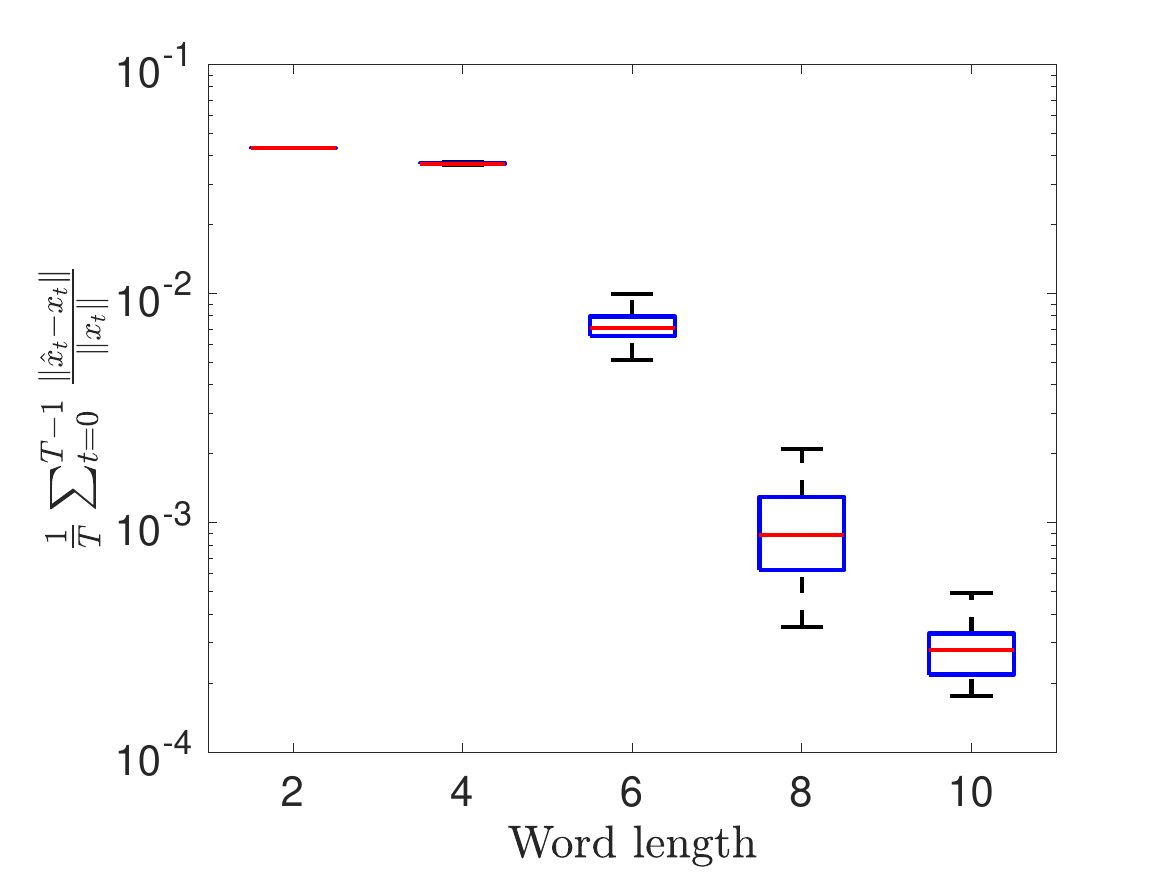}}
\subfloat[]{\includegraphics[trim=0cm 0cm 0cm 0cm, clip=true, width=0.33\textwidth]{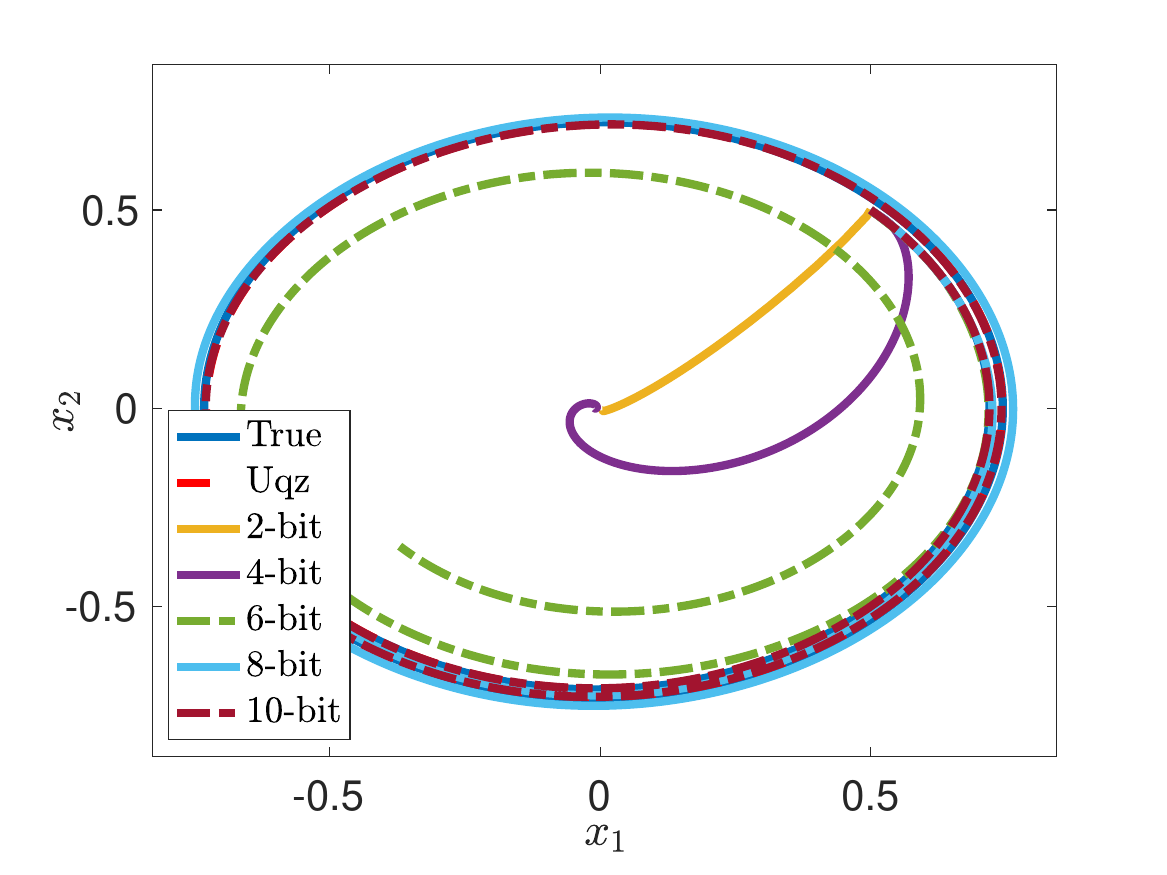}}
\caption{Error and prediction profile for negatively-damped pendulum \eqref{Eq: Pendulum}.} \label{Fig: Pendulum}
\end{figure*}

The relative 2-norm error $\frac{\| \Kd - \Kdt\|}{\|\Kd\|}$ for EDMD matrix, time-average relative two norm error $\frac{1}{T}\sum_{t=0}^{T-1} \frac{\|\hat{x}_t - x_t \|}{\|x_t\|}$ of predictions using $\Kdt$ for different word-length, and corresponding predicted trajectories in phase-space are shown in Fig.~\ref{Fig: Pendulum}. For the first two plots, a Monte-Carlo run for 50 independent dither signals are used. 

We notice that the prediction error in Fig.~\ref{Fig: Pendulum}(b) decreases with the quantization word length, as one would expect.
The trend is consistent in first two subplots, where the median relative errors (shown by the red line segments) decrease with the word length. 
In Fig.~\ref{Fig: Pendulum}(c) we plot the actual system trajectories as the quantization word length is varied.  
This experiment demonstrates that a word length of 8 is enough to obtain a satisfactory system identification.  
\begin{figure*}[t]
\centering 
\subfloat[]{\includegraphics[trim=0cm 0cm 0cm 0cm, clip=true, width=0.33\textwidth]{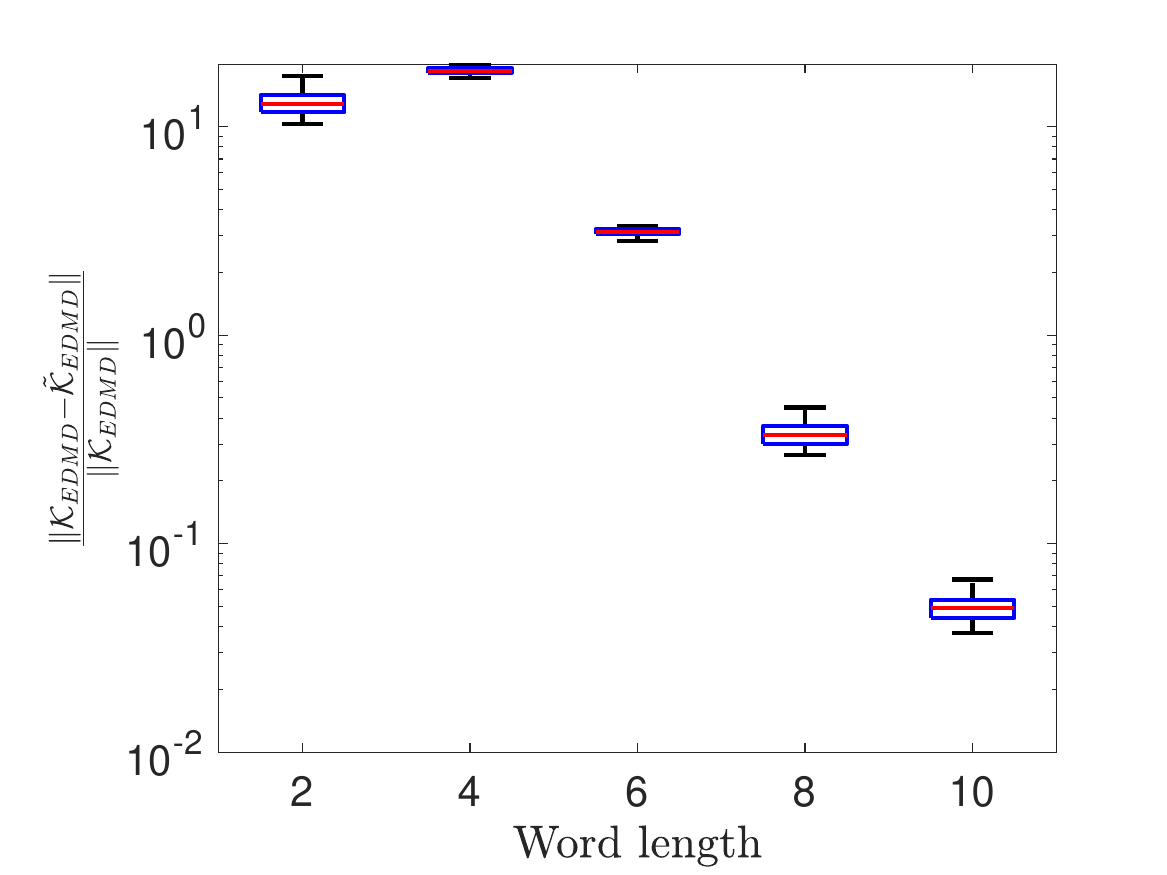}}
\subfloat[]{\includegraphics[trim=0cm 0cm 0cm 0cm, clip=true, width=0.33\textwidth]{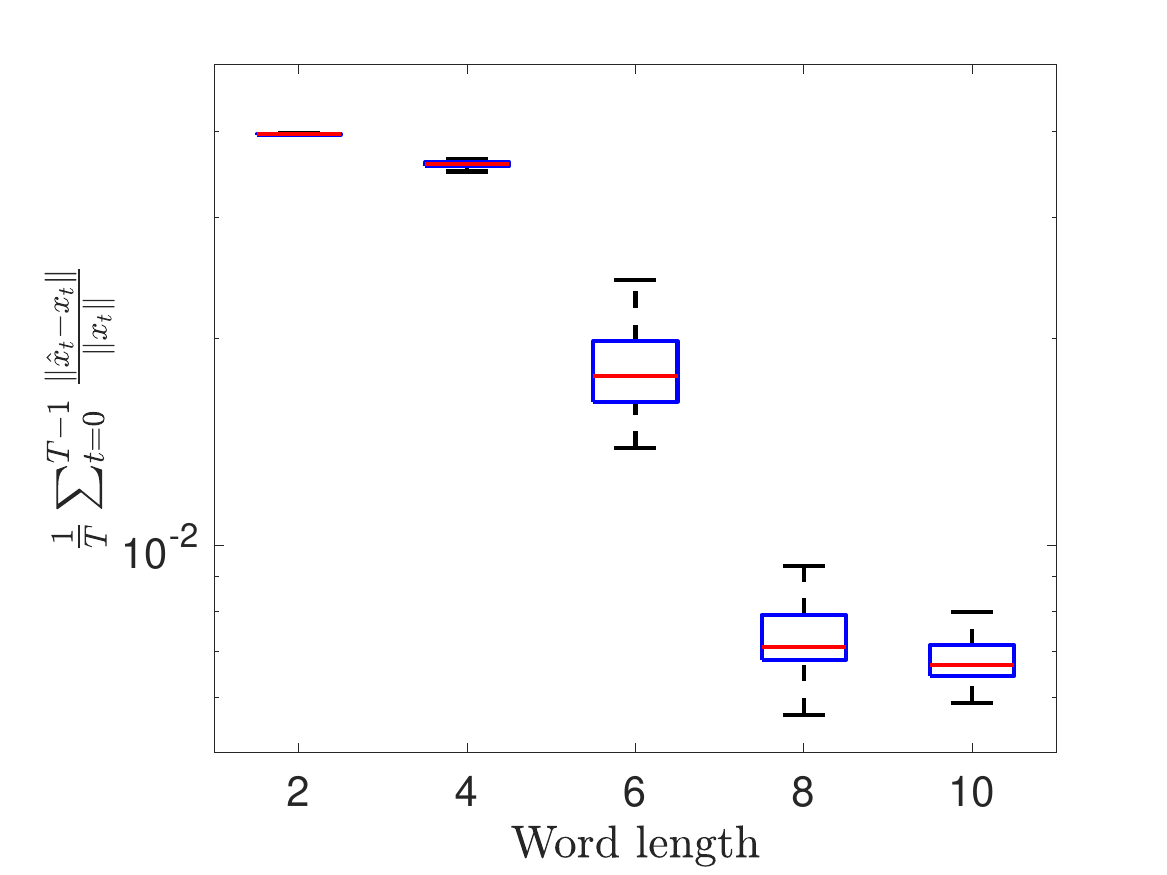}}
\subfloat[]{\includegraphics[trim=0cm 0cm 0cm 0cm, clip=true, width=0.33\textwidth]{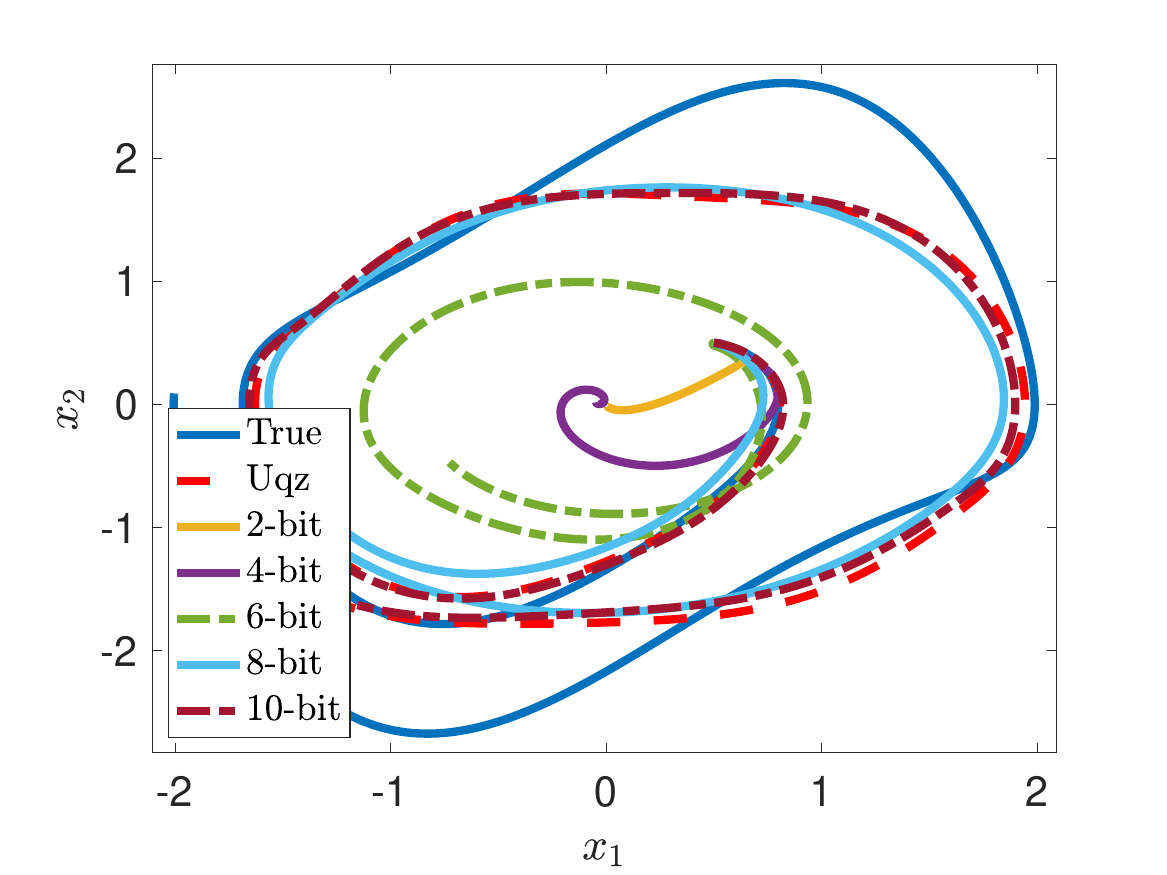}}
\caption{Error and prediction profile for Van der Pol oscilator \eqref{Eq: Vanderpol}.} \label{Fig: VDP}
\end{figure*}\vspace{-0.3cm}

\subsection{Van der Pol oscillator}
Now, we consider the limit-cyclic Van der Pol oscillator:
\bnl \label{Eq: Vanderpol}
\dot{x}_1 &=& x_2\nonumber\\
\dot{x}_2&=&(1-x_1^2)x_2-x_1,
\enl
with the same $\Delta t$, integration-scheme, and number of samples. The initial conditions are generated randomly with uniform distribution on the box $[-2, 2]^2$. The lifting functions $\varphi^i$ are chosen to be the state itself (i.e., $\varphi^1=x_1$, $\varphi^2=x_2$) and 100 thin plate spline radial basis functions with centers selected randomly with uniform distribution on the $[-2,2]^2$ box. The dimension of the lifted state-space is therefore $N = 102$ here as well. In this experiment too, we notice the same trend in Fig.~\ref{Fig: VDP}. 
Fig.~\ref{Fig: VDP}(c) plots the predicted trajectories with different levels of quatization. A word-length of 8-bit is sufficient here as well for system-identification performance comparable to unquantized data.

\section{Conclusions} \label{sec:conclusions}
In this letter, we present the $\DQEDMD$---a least-square optimization method that estimates the Koopman operator from dither quantized data. 
We theoretically analyze the connection between the estimates obtained from the quantized data and that from unquantized data. 
The effect of quantization is analyzed and quantified for both finite and large data regimes. 
Our analysis shows the quantization resolution $\epsilon$ affects the estimates as $O(\epsilon)$ in finite data regime and $O(\epsilon^2)$ in large data regime. 
The analysis is validated via repeated trials of experiments on multiple problems.

\bibliographystyle{ieeetr}
\bibliography{references, ref1}

\appendix

\subsection{Proof of \Cref{thm:equivalence}} \label{AP:thm:equivalence} 
\noindent Notice that we may express $\| \bar \Phi' - A \bar \Phi\|^2$ as 
\begin{align} \label{eq:phi_sum}
     \| \bar \Phi' - A \bar \Phi\|^2 &=  \sum\nolimits_{t=0}^{T-1}\| \bar \varphi(x_{t+1}) - A \bar \varphi(x_t)\|^2\\\nonumber
     &= \sum\nolimits_{t=0}^{T-1}\|  \varphi(x_{t+1}+e_{t+1}) - A  \varphi(x_t+e_t)\|^2 \\\nonumber
     &= \sum\nolimits_{t=0}^{T-1}r(x_{t+1}+e_{t+1},x_t+e_t).
\end{align}
Expanding $r(x_{t+1}+e_{t+1},x_t+e_t)$ via Taylor series we get
\begin{align}\label{eq:r_sum}
    & r(x_{t+1}\!+\!e_{t+1},x_t\!+\!e_t) = r(x_{t+1},x_t) + \lim_{n\rightarrow\infty}\sum\nolimits_{k=1}^n h_k(e_{t+1},e_t), 
\end{align}
where, the $k$-th term $h_k(\cdot,\cdot)$ involves the $k$-th order derivative. For instance,
\begin{align*}
    h_1(e_{t+1},e_t) &= \nabla_{x_{t+1}}r(x_{t+1},x_t)^\top e_{t+1} + \nabla_{x_{t}}r(x_{t+1},x_t)^\top e_{t}, \\
    h_2(e_{t+1}, e_t) &= \dfrac{1}{2} e_{t+1}^\top \nabla^2_{x_{t+1}}r(x_{t+1},x_t) e_{t+1} + \dfrac{1}{2} e_{t}^\top \nabla^2_{x_{t}}r(x_{t+1},x_t) e_{t} \\
     & + e_{t+1}^\top \nabla_{x_{t+1}}\nabla_{x_{t}}r(x_{t+1},x_t) e_{t}.
\end{align*}
For $T\rightarrow\infty$, we may write
\begin{align}\label{eq:residual_large_data}
    &\lim_{T\rightarrow\infty}\dfrac{1}{T}\sum\nolimits_{t=0}^{T-1} r(x_{t+1}  + e_{t+1},  x_t+e_t) = \\
    & \lim_{T\rightarrow\infty}\dfrac{1}{T}\sum\nolimits_{t=0}^{T-1} r(x_{t+1}, x_t) \nonumber  +  \lim_{n\rightarrow\infty}\sum\limits_{k=1}^n \lim_{T\rightarrow\infty} \dfrac{1}{T}\sum\nolimits_{t=0}^{T-1} h_k(e_{t+1},e_t), 
\end{align}
where we have used the Assumption~\ref{assm:AbsConvTaylor} and invoked Fubini's theorem to interchange the order of summation in the last term of \eqref{eq:residual_large_data}. 
Next, we will simplify each term {\small $\lim_{T\rightarrow\infty} \dfrac{1}{T}\sum\nolimits_{t=0}^{T-1} h_k(e_{t+1},e_t)$} using Kolmogorov's strong law of large numbers. 
For $k=1$:
\begin{align*}
    \lim_{T\rightarrow\infty} \dfrac{1}{T}\sum\nolimits_{t=0}^{T-1} h_1(e_{t+1},e_t) & = \lim_{T\rightarrow\infty} \dfrac{1}{T}\sum\nolimits_{t=0}^{T-1} \nabla_{x_{t+1}}r(x_{t+1},x_t)^\top e_{t+1} \\
    &\qquad + \lim_{T\rightarrow\infty} \dfrac{1}{T}\sum\nolimits_{t=0}^{T-1} \nabla_{x_{t}}r(x_{t+1},x_t)^\top e_{t} \\
   & \overset{\text{almost surely}}{\longrightarrow} 0, 
\end{align*}
where we have used Kolmogorov's strong law of large numbers: {\small$\lim_{T\rightarrow\infty} \dfrac{1}{T}\sum\nolimits_{t=0}^{T-1} \nabla_{x_{t+1}}r(x_{t+1},x_t)^\top e_{t+1} \to \lim_{T\rightarrow\infty}  \nabla_{x_{t+1}}r(x_{t+1}, x_t)^\top \E[e_{t+1}]  = 0$} almost surely.\footnote{Applying law of large number requires $\nabla_{x_{t+1}}r(x_{t+1},x_t)^\top e_{t+1}$ to have a finite second moment and $\sum_{t=1}^\infty \frac{1}{t^2} \text{Var}(\nabla_{x_{t+1}}r(x_{t+1},x_t)^\top e_{t+1})$ to be finite.
Both of these conditions are satisfied due to Assumptions~\ref{assm:boundedR}--\ref{assm:BoundedPhi_derivative}. } 
Similarly, we simplify the $k=2$ term: 
\begin{align*}
    \lim_{T\rightarrow\infty} \dfrac{1}{T}\!\!\sum\nolimits_{t=0}^{T-1} h_2(e_{t+1},e_t) & = \lim_{T\rightarrow\infty} \dfrac{1}{T}\!\!\sum\nolimits_{t=0}^{T-1}  \dfrac{1}{2} e_{t+1}^\top \nabla^2_{x_{t+1}}r(x_{t+1},x_t) e_{t+1} \\
    &\quad + \lim_{T\rightarrow\infty} \dfrac{1}{T}\!\!\sum\nolimits_{t=0}^{T-1} \dfrac{1}{2} e_{t}^\top \nabla^2_{x_{t}}r(x_{t+1},x_t) e_{t} \\
    & + \lim_{T\rightarrow\infty} \dfrac{1}{T}\!\!\sum\nolimits_{t=0}^{T-1} e_{t+1}^\top \nabla_{x_{t+1}}\nabla_{x_{t}}r(x_{t+1},x_t) e_{t}.
\end{align*}
 Using Kolmogorov's law of large numbers, we notice that 
 \begin{align*}
     \dfrac{1}{T}\sum\limits_{t=0}^{T-1}  \dfrac{1}{2} e_{t+1}^\top \nabla^2_{x_{t+1}}r(x_{t+1},x_t) e_{t+1} \to \frac{\epsilon^2}{24} \dfrac{1}{T}\sum\limits_{t=0}^{T-1}  \tr(\nabla^2_{x_{t+1}}r(x_{t+1},x_t)),
 \end{align*}
 where we have used $\E[e_te_t^\top] = \frac{\epsilon^2}{12}I$ for all $t$. 
 Similarly, \vspace{-0.25cm}
  \begin{align*}
     &\dfrac{1}{T}\sum\nolimits_{t=0}^{T-1}  e_{t}^\top \nabla^2_{x_{t}}r(x_{t+1},x_t) e_{t} \to \frac{\epsilon^2}{24} \dfrac{1}{T}\sum\nolimits_{t=0}^{T-1}  \tr(\nabla^2_{x_{t}}r(x_{t+1},x_t)), \\
     & \dfrac{1}{T}\sum\nolimits_{t=0}^{T-1} e_{t+1}^\top \nabla_{x_{t+1}}\nabla_{x_{t}}r(x_{t+1},x_t) e_{t} \to 0
 \end{align*}
 where the last result is obtained by using $\E[e_{t+1}e_t^\top] = 0$. More details can be found in the Appendix of \cite{maity2024effect}.
 Now, notice that \vspace{-0.1cm}
 \begin{align*}
     \nabla^2_{x_{t+1}}r(x_{t+1},x_t) &= (\nabla_{x_{t+1}}\varphi(x_{t+1}))(\nabla_{x_{t+1}}\varphi(x_{t+1}))^\top \\
      &+ \sum_{i=1}^N  [\varphi(x_{t+1}) - A \varphi(x_t)]_i \nabla^2_{x_{t+1}} \varphi^i(x_{t+1}),
 \end{align*}
where $[\varphi(x_{t+1}) - A \varphi(x_t)]_i$ is the $i$-th component of the vector. 
Consequently, 
\begin{align*}
    \tr(\nabla^2_{x_{t+1}}r(x_{t+1},x_t)) &= \|\nabla_{x_{t+1}}\varphi(x_{t+1})\|^2 \\
    & + (\varphi(x_{t+1}) - A \varphi(x_t))^\top g(x_{t+1}),
\end{align*}
where {\small$g(x_{t+1})\in \R^N$} with {\small$\tr(\nabla^2_{t+1} \varphi^i(x_{t+1}))$} being its $i$-th component.
Similarly, we obtain 
\begin{align*}
    \tr(\nabla^2_{x_{t}}r(x_{t+1},x_t)) &= \|A\nabla_{x_{t}}\varphi(x_{t})\|^2  - (\varphi(x_{t+1}) - A \varphi(x_t))^\top A g(x_t).
\end{align*}
Therefore, we may write 
\begin{align*}
    \dfrac{1}{T}\sum\nolimits_{t=0}^{T-1} h_2(e_{t+1},e_t) \to \epsilon^2\left(\alpha_2 + \tr(A\beta_2) + \tr(A^\top A\Gamma_2)  \right),
\end{align*}
where \vspace{-0.3cm}
\begin{align*}
    \alpha_2 &= \lim_{T\to \infty}\dfrac{1}{24T}\sum\nolimits_{t=0}^{T-1} \|\nabla_{x_{t+1}}\varphi(x_{t+1})\|^2 + \varphi(x_{t+1})^\top g(x_{t+1}) \\
    \beta_2 &= - \lim_{T\to \infty}\dfrac{1}{24T}\sum\nolimits_{t=0}^{T-1} 2\varphi(x_t)g(x_{t+1})^\top,\\
    \Gamma_2 &= \lim_{T\to \infty}\dfrac{1}{24T}\sum\nolimits_{t=0}^{T-1} \nabla_{x_{t}}\varphi(x_{t}) \nabla_{x_{t}}\varphi(x_{t})^\top + g(x_t) \varphi(x_t)^\top.
\end{align*}
Similarly, one may verify that for any $k$,
\begin{align}
     \dfrac{1}{T}\!\!\sum\limits_{t=0}^{T-1}\! h_k(e_{t+1},e_t) \overset{\text{almost}}{\underset{\text{surely}}{\to}}  \begin{cases}
         \epsilon^k\left(\alpha_k + \tr(A\beta_k) + \tr(A^\top A\Gamma_k)  \right), ~ k = \text{even} \\
         0,\qquad \qquad  \text{otherwise}.
     \end{cases}
\end{align}
Adding all the residuals, 
\begin{align}\label{eq:residuals}
    \lim_{T\rightarrow\infty}\dfrac{1}{T}\!\!\sum\nolimits_{t=0}^{T-1} r(x_{t+1} & + e_{t+1}, x_t+e_t) \to \lim_{T\rightarrow\infty}\dfrac{1}{T}\!\!\sum\nolimits_{t=0}^{T-1} r(x_{t+1}, x_t) \nonumber \\
    &+ \alpha(\epsilon) + \tr(A \beta(\epsilon)) + \tr(A\T A \Gamma(\epsilon))  ,    
\end{align}
where {\small $\alpha(\epsilon) = \sum_{k=1}^\infty \epsilon^{2k} \alpha_{2k}$, $\beta(\epsilon) = \sum_{k=1}^\infty \epsilon^{2k} \beta_{2k}$}, and {\small $\Gamma(\epsilon) = \sum_{k=1}^\infty \epsilon^{2k} \Gamma_{2k}$}.

\noindent
Consequently, from \eqref{eq:phi_sum} and \eqref{eq:residuals}
   \begin{align} 
    \begin{split}
        \Kdt &= \argmin_{A \in \R^{n\times n}} \frac{1}{T} \| \bar \Phi' - A \bar \Phi\|^2  \\ 
        &\underset{T\rightarrow\infty}{\overset{\operatorname{a.s.}}{\longrightarrow}} \limsup_{T\rightarrow\infty}\argmin_{A \in \mathcal{A}} \frac{1}{T}  \| \Phi' - A \Phi\|^2 \\
        &\qquad + \tr(A\beta(\epsilon)) + \tr(A\T A \Gamma(\epsilon)).
        \end{split}
    \end{align}
This completes the proof. \hfill $\blacksquare $

\end{document}